\documentclass[prl,nofootinbib]{revtex4}
\usepackage{graphicx}
\usepackage{latexsym}
\def\be{\begin{equation}}
\def\ee{\end{equation}}
\def\bea{\begin{eqnarray}}
\def\eea{\end{eqnarray}}

\newcommand{\DE}{{\mathrm{DE}}}
\newcommand{\tot}{{\mathrm{tot}}}
\newcommand{\eff}{{\mathrm{eff}}}

\begin{document}

\title{Constraints on the Sound Speed of Dynamical Dark Energy}

\author{Jun-Qing Xia}
\author{Yi-Fu Cai}
\author{Tao-Tao Qiu}
\author{Gong-Bo Zhao}
\author{Xinmin Zhang}

\affiliation{Institute of High Energy Physics, Chinese Academy of
Science, P.O. Box 918-4, Beijing 100049, P. R. China}


\begin{abstract}

In this paper we study the sound speed - $c_s^2$, which is
directly related to the classical perturbations - of the dynamical
dark energy (DE), especially with an equation of state crossing
the cosmological constant boundary in details and show its
implications on Cosmic Microwave Background (CMB) Anisotropy. With
the present observational data of CMB, Type Ia Supernova (SNIa)
and galaxy clustering, we perform a global analysis to constrain
the sound speed of dark energy, using the Markov Chain Monte Carlo
method. We find that the sound speed of dark energy is weakly
constrained by current observations thus the futuristic precision
measurements of CMB on a very large angular scale (low multipoles)
are necessary.

\end{abstract}


\maketitle

\section{I. Introduction}

The analysis of the redshift-distance relation of SNIa revealed
that our Universe is currently accelerating \cite{Riess98,Perl99}
which has been confirmed by the current high quality observations
of the CMB \cite{wmap3:2006}, the Large Scale Structure (LSS) of
galaxies
 \cite{Cole:2005sx,Tegmark:2003uf,Tegmark:2003ud,Tegmark:2006az},
and the SNIa \cite{Tonry03,Riess04,Riess05,Riess:2006fw}. One
possible explanation for this phenomenon is that this acceleration
is attributed to a new form of energy, dubbed \emph{Dark Energy},
which recently dominate the energy density of the Universe with
negative pressure and almost not clustering. The nature of DE is
among the biggest problems in modern physics and has been studied
extensively. A cosmological constant, the simplest DE candidate
whose equation of state $w$ remains $-1$, suffers from the
well-known fine-tuning and coincidence problems. Alternatively,
dynamical dark energy models with rolling scalar fields have been
proposed, such as Quintessence \cite{quint1,quint2,pquint},
Phantom \cite{phantom}, Quintom \cite{Feng:2004ad} and K-essence
\cite{Chiba:1999ka,kessence1,kessence2}. Given our ignorance of
the nature of dark energy, the cosmological observations play a
crucial role in our understanding of DE. There are many studies on
DE both theoretically and phenomenologically in the literature
\cite{Feng:2004ff,xia,Zhao:2005vj,Xia:2005ge,
Xia:2006cr,Zhao:2006bt,Xia:2006rr,Li:2005fm,Zhang:2005eg,Zhang:2006ck,
Guo:2004fq,Copeland:2006wr,Sahni:2002dx,Alam:2003fg,Alam:2006kj,Wei:2005nw,Cai:2005ie,Andrianov:2005tm,
Zhang:2005yz,Guo:2005nu,McInnes:2005vp,Aref'eva:2005fu,Vikman,Huang:2005gu,
Zhao:2006mp,Grande:2006nn,Chimento:2006xu,Cannata:2006gd,Lazkoz:2006pa,
Stefancic:2005nu,Perivolaropoulos:2006ce}.

If we regard the mysterious DE as a cosmic fluid, there will be
perturbations when $w\neq-1$. In the framework of the linear
perturbations theory, the DE perturbations can be fully
characterized by its equation of state, $w\equiv p/\rho$, and its
sound speed, $c_s^2\equiv \delta p/ \delta \rho$. For the DE
perturbations it is well-defined in the region of Quintessence,
$w>-1$, and Phantom, $w<-1$. In our previous works
\cite{Zhao:2005vj,Xia:2005ge} based on Quintom dark energy model,
we have proposed a new method to handle the DE perturbations in
the region closed to $w=-1$, assuming the sound speed of Dark
Energy is equal to unity, $c_s^2=1$. We found that the constraints
on the DE parameters are relaxed and the parameter space are
enlarged dramatically when the DE perturbations are included in
the analysis. The effects of perturbations on studying the feature
of DE are of great significance.

Since the sound speed of DE merely affects the evolution of
perturbation, it has not been investigated extensively in the
literature. Recently the observational implications of the sound
speed on CMB and LSS of some DE models have been discussed: for
example, for K-essence \cite{Erikson01,Dedeo03}, condensation of
Dark Matter \cite{Bassett02} and the Chaplygin gas \cite{
Sandvik:2002jz,Beca:2003an,Reis:2003mw,Bean:2003ae,Amendola:2003bz}.
In Refs.\cite{Weller:2003hw,Bean:2003fb,hannestad} the authors
tried to use the current observational data to constrain the sound
speed of DE. However, they just considered the equation of state
of DE to be constant. In this paper we consider the dynamical DE
model with time varying $w$ and extend our previous works on the
sound speed of DE models and allow an arbitrary constant sound
speed in our calculation. Combining the present observational
data, such as CMB, LSS and SNIa, we discuss the possible
constraints on the sound speed of DE. Our paper is organized as
follows: In Section II we analyze the feature of the sound speed
of dark energy models paying attention to the case when the
equation of state gets across $w=-1$; In Section III we describe
the method and the datasets we used; In Section IV we present our
results derived from a global analysis using Markov Chain Monte
Carlo (MCMC) method; Finally, Section V contains a discussion of
the results.

\section{II. Feature of Sound Speed}
\subsection{A. Single Perfect Fluid}

Working in the conformal Newtonian gauge, one can easily describe
the DE perturbations as follows \cite{ma}:
\begin{eqnarray}
    \delta'&=&-(1+w)(\theta-3\Phi')
    -3\mathcal{H}(c_{s}^2-w_i)\delta~, \label{dotdelta}\\
\theta'&=&-\mathcal{H}(1-3w)\theta-\frac{w'}{1+w}\theta
    +k^{2}(\frac{c_{s}^2\delta}{{1+w}}+ \Psi)~ , \label{dottheta}
\end{eqnarray}
where the prime denotes the derivative with respect to conformal
time, $c_{s}^2\equiv \delta p/ \delta \rho$ denotes the sound
speed of DE models, $\delta$ and $\theta$ are the density and
velocity perturbations of DE models respectively, $w\equiv p/\rho$
is the equation of state, and $\Phi$ and $\Psi$ represent the
metric perturbations.


Let us discuss the behavior of the perturbations of single perfect
fluid when $w$ crosses the boundary $w=-1$ in the conformal
Newtonian gauge, following \cite{Bean:2003fb}. Firstly, we
consider the DE as a single barotropic fluid. The adiabatic speed
of sound, $c_{a}^{2}$, is purely determined by the equation of
state $w$,
\begin{eqnarray}
\delta p=\left(\frac{p'}{\rho'}\right)\delta\rho\equiv
c_{a}^{2}\delta\rho=\left(w-{w'\over 3{\cal H}(1+w)}\right)\delta
\rho ~.\label{adisoseq}
\end{eqnarray}
Whenever the equation of state closes to the boundary $w=-1$ and
$w'|_{w=-1}\neq0$, the adiabatic speed of sound, $c_{a}^{2}$, will
be divergent due to the existence of the term $w'/(1+w)$.
Apparently, the perturbations of this system become unstable.

However, in non-barotropic fluids, for example in most scalar
field models, this simple relation in Eq.(\ref{adisoseq}) between
the equation of state and the sound speed breaks down, because of
the intrinsic entropy perturbations, and we have the more general
relation
\begin{eqnarray}\label{cs2}
c_{s}^{2}&\equiv&{\delta p\over \delta\rho}~.
\end{eqnarray}
In this case, if $c_{s}^{2}\neq c_{a}^2$, the intrinsic entropy
perturbation $\Gamma$ will be induced \cite{Bean:2003fb,Kodama84}:
\begin{eqnarray}
w\Gamma  \equiv  (c_{s}^{2} - c_{a}^2)\delta =
{p'\over\rho}\left({\delta p\over p'}-{\delta \rho\over
\rho'}\right)~. \label{Gammaeq}
\end{eqnarray}
Whereas the adiabatic sound speed, $c_{a}^{2}$, and $\Gamma$ are
scale independent and gauge invariant quantities, while
$c_{s}^{2}$ can be neither. In order to construct the gauge
invariant $c_{s}^{2}$ we use a helpful transformation
\cite{Kodama84}:
\begin{eqnarray}
\hat{\delta}=\delta+3{\cal H}(1+w){\theta\over
k^{2}}~.\label{Deltaeq}
\end{eqnarray}
This transformation relates the gauge invariant, rest frame
density perturbation, $\hat{\delta}$, to the density and velocity
perturbations in a general frame, $\delta$ and $\theta$. Using
Eqs.(\ref{Gammaeq},\ref{Deltaeq}), we can rewrite the pressure
perturbation in a general frame, $\delta p$,  in terms of the rest
frame sound speed, $\hat{c}_{s}^{2}$,
\begin{eqnarray}
\delta p &=&\hat{c}_{s}^{2}\delta\rho+3{\cal
  H}(1+w)(\hat{c}_{s}^{2}-c_{a}^{2})\rho{\theta\over
  k^{2}}\label{deltapeq}~.
\end{eqnarray}
For example single scalar field models with standard kinetic terms
always have $\hat{c}_{s}^{2}=1$. Moreover, the
Eqs.(\ref{dotdelta},\ref{dottheta}) can be rewritten as:
\begin{eqnarray}
\delta'&=&-(1+w)(\theta-3\Phi')
-3\mathcal{H}(\hat{c}_{s}^2-w)\delta-3\mathcal{H}(w'+3\mathcal{H}(1+w)(\hat{c}_{s}^2-w))\frac{\theta}{k^2}~, \label{dotdelta2}\\
\theta'&=&-\mathcal{H}(1-3\hat{c}_{s}^2)\theta+k^{2}(\frac{\hat{c}_{s}^2\delta}{{1+w}}+
\Psi)~ . \label{dottheta2}
\end{eqnarray}
In this case if we require the DE equation of state can cross the
boundary $w=-1$, the DE perturbations are still divergent (see
appendix for details).

\subsection{B. Two-field Quintom Dark Energy Models}

As discussed above, the DE models, whose equation of state can
cross the boundary $w=-1$, need more degrees of freedom to keep
the whole system stable \cite{Zhao:2005vj,Hu:2004kh}. As an
example, we consider one type of Quintom DE models with two
components, one is Quintessence-like, $-1\leq w\leq1$, and the
other is Phantom-like, $w\leq-1$
\cite{Feng:2004ad,Zhao:2005vj,Kunz:2006wc} and analyze the
perturbations of this DE system.

The most notable advantage, in this case, is that it is
unnecessary to let the total equation of state, $w_\tot$, of this
Quintom system cross the boundary $w=-1$ by forcing the equation
of state of each component to cross the boundary. For the
Quintessence-like component, $-1\leq w\leq1$, while for the
Phantom-like component, $w\leq-1$, during the evolution, without
any crossing behavior. For each component of the Quintom model the
density and velocity perturbations, $\delta_{i}$ and $\theta_{i}$,
still satisfy the Eqs.(\ref{dotdelta2},\ref{dottheta2}). Because
the equation of state for each component needs not cross the
boundary, the perturbations of each component will be stable and
all the variables will be well defined. Furthermore, all these
corresponding variables of the whole Quintom DE model can be
combined with each component:
\begin{eqnarray}
\bar{\rho}_\tot&=& \bar{\rho}_{1} + \bar{\rho}_{2}~, \label{rhoeff}\\
\bar{p}_\tot&=& \bar{p}_{1} + \bar{p}_{2} \label{peff}~,\\
w_\tot&=& \frac{w_{1}\bar{\rho}_{1} + w_{2}\bar{\rho}_{2}}{\bar{\rho}_{1}+\bar{\rho}_{2}}~\label{weff},\\
\delta_\tot&=& \frac{\bar{\rho}_{1}\delta_{1} + \bar{\rho}_{2}\delta_{2}}{\bar{\rho}_{1}+\bar{\rho}_{2}}~, \label{deltaeff}\\
\delta{p}_\tot&=& \delta{p}_{1} + \delta{p}_{2}
\label{deltapeff}~,\\
\theta_\tot&=&\frac{\left( 1+w_{1}
\right)\bar{\rho}_{1}\theta_{1}+\left( 1+w_{2}
\right)\bar{\rho}_{2}\theta_{2}}{\left( 1+w_{1}
\right)\bar{\rho}_{1}+\left( 1+w_{2} \right)\bar{\rho}_{2}}
\label{thetaeff}~.
\end{eqnarray}
Using these Eqs.(\ref{rhoeff}$-$\ref{deltapeff}), these variables
of Quintom DE models are well defined except the velocity
perturbations $\theta_\tot$.

As long as $w_\tot$ gets across the boundary,
$w_\tot\rightarrow-1$, the denominator of Eqs.(\ref{thetaeff})
goes to zero, $[\left( 1+w_{1} \right)\bar{\rho}_{1}+\left(
1+w_{2} \right) \bar{\rho}_{2}]\rightarrow0$. Consequently, the
total velocity perturbations, $\theta_\tot$, become divergent
unless these two independent perturbations are equal all the time,
$\theta_\tot=\theta_1=\theta_2$. During the evolution, it is
impossible to keep the velocity perturbations of each component
equal, $\theta_1=\theta_2$, all the time. However, the physically
meaningful velocity perturbation which is relevant to the CMB
observations is $\mathcal{V}\equiv(1+w)\theta$, not $\theta$ alone
\cite{Zhao:2005vj,Caldwell:2005ai}. Eq.(\ref{thetaeff}) can now be
written as:
\begin{equation}
\mathcal{V}_\tot=\frac{\mathcal{V}_{1}\bar{\rho}_{1} +
\mathcal{V}_{2}\bar{\rho}_{2}}{\bar{\rho}_{1}+\bar{\rho}_{2}}
\label{Veff}~.
\end{equation}
Using this Eq.(\ref{Veff}), the divergences disappear when the
total equation of state crosses the boundary $w=-1$. And then all
the variables are well defined and the perturbations of Quintom
dark energy system are stable in the whole parameter space of the
total equation of state, adding more degrees of freedom.

Let us move to the discussion of the sound speed of the whole
Quintom dark energy system. Physically we can use the independent
sound speed of each component to describe the whole system.
However, the present constraints on the sound speed of DE are so
weak that considering to study the two independent sound speed is
not justified at present. For simplicity we investigate the
effective sound speed, $\hat{c}_{s,\eff}^{2}$ \cite{Kodama84}:
\begin{equation}
\hat{c}_{s,\eff}^{2}=\frac{\hat{c}_{s,1}^{2}\left( 1+w_{1} \right)
\bar{\rho}_{1}+ \hat{c}_{s,2}^{2}\left( 1+w_{2} \right)
\bar{\rho}_{2}}{\left( 1+w_{1} \right) \bar{\rho}_{1}+\left(
1+w_{2} \right) \bar{\rho}_{2}} \label{cs2eff}~.
\end{equation}
It is apparent that this effective sound speed suffers from the
same problem with the velocity perturbations $\theta_\tot$. In
order to settle this divergence of the effective sound speed we
fix the sound speed of each component to be equal,
$\hat{c}_{s,1}^2=\hat{c}_{s,2}^2=\hat{c}_{s}^2$. The effective
sound speed of the whole Quintom system should be a constant
during the evolution:
\begin{equation}
\hat{c}_{s,\eff}^{2}=\frac{\hat{c}_{s}^{2}\left( 1+w_{1} \right)
\bar{\rho}_{1}+ \hat{c}_{s}^{2}\left( 1+w_{2} \right)
\bar{\rho}_{2}}{\left( 1+w_{1} \right) \bar{\rho}_{1}+\left(
1+w_{2} \right)
\bar{\rho}_{2}}=\hat{c}_{s}^{2}=\hat{c}_{s,1}^2=\hat{c}_{s,2}^2
\label{constcs2}~.
\end{equation}
The total equation of state of Quintom system has nothing to do
with the effective sound speed now.

In our previous works \cite{Zhao:2005vj,Xia:2005ge} we considered
the Quintom dark energy model with the effective sound speed
$\hat{c}_{s,\eff}^{2}=1$. In this paper we relax this limitation
and assume an effective arbitrary constant sound speed of Quintom
DE model, $\hat{c}_{s,\eff}^{2}=\hat{c}_{s}^{2}$. The effects of
this effective sound speed on CMB power spectrum will be studied
in the next subsection.

\subsection{C. Implications on CMB Anisotropy}

\begin{figure}[htbp]
\begin{center}
\includegraphics[scale=0.6]{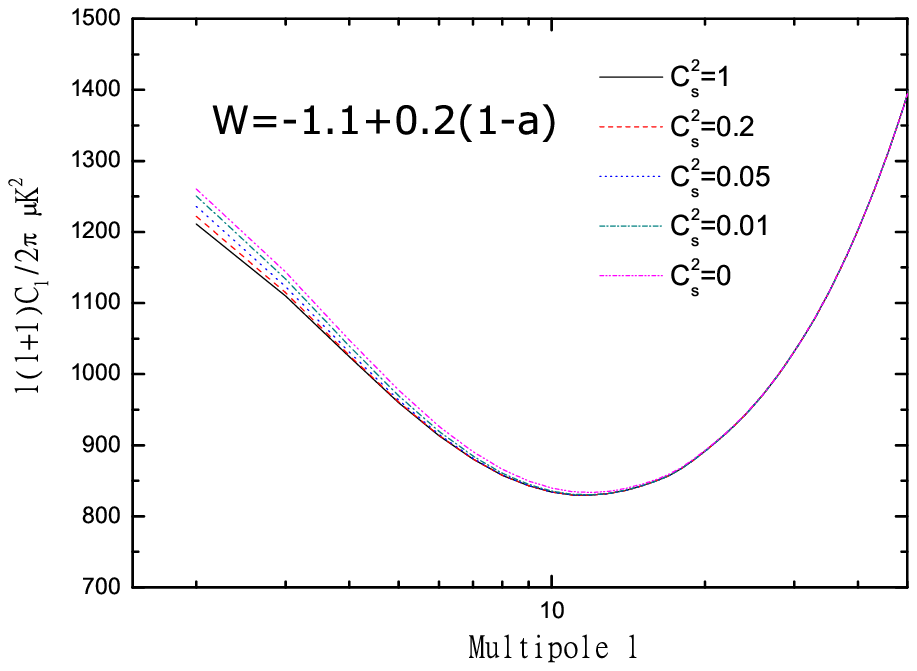}
\includegraphics[scale=0.6]{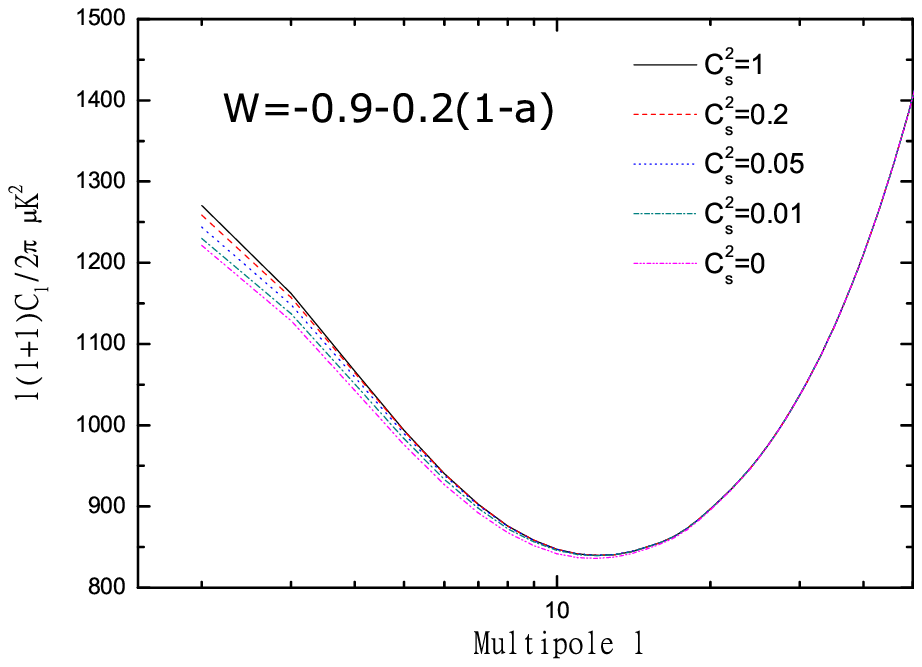}
\includegraphics[scale=0.6]{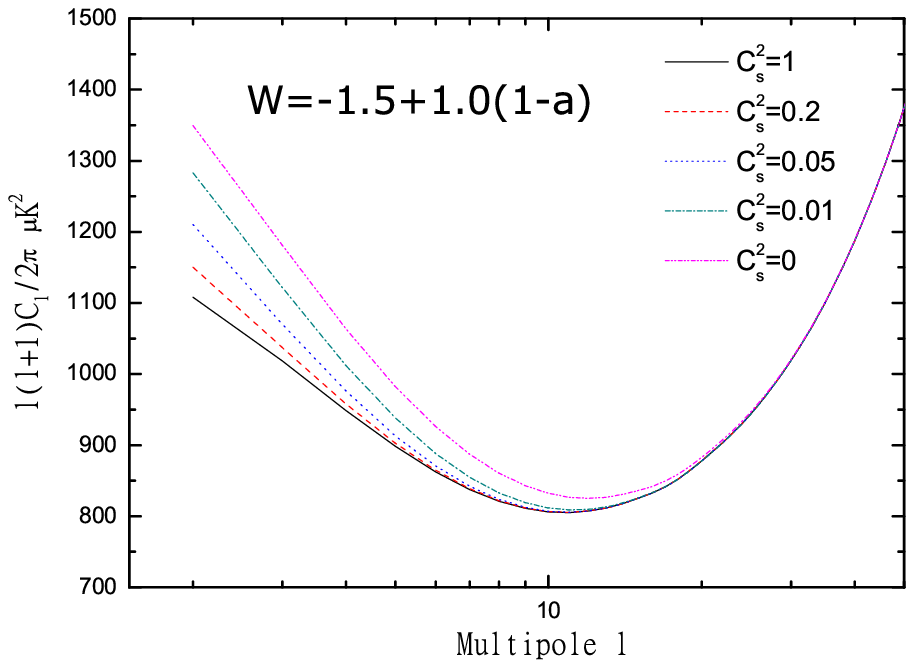}
\includegraphics[scale=0.6]{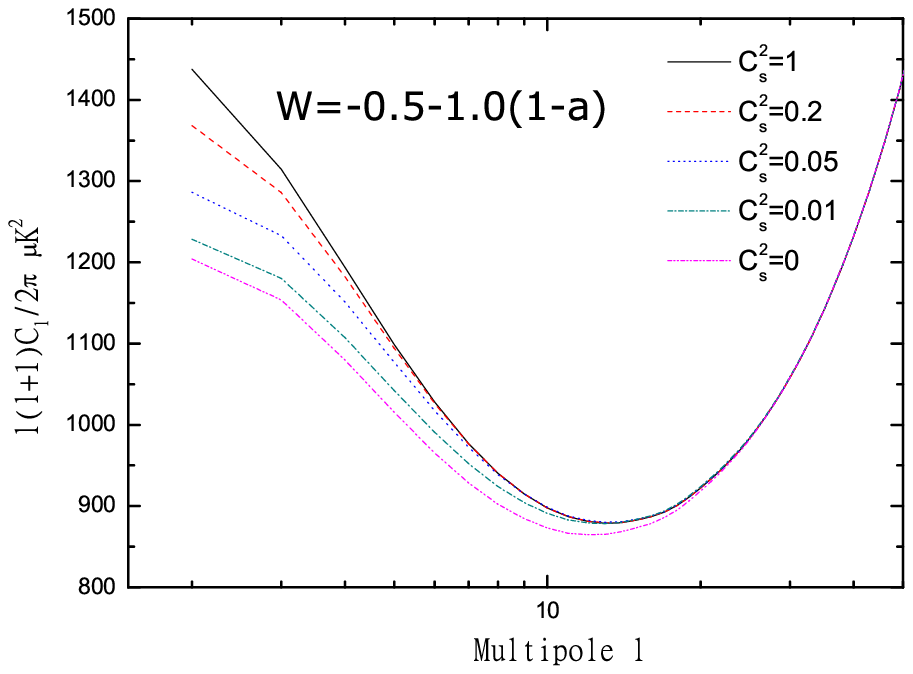}
\caption{The effects on CMB temperature power spectra of four
different Dark Energy models charactrized by different equation of
states $w$, for different constant Dark Energy sound speed
$\hat{c}_{s}^{2}$. On the left two panels, the bottom Black solid
line is for the sound speed of $\hat{c}_{s}^{2}=1.0$ and the top
Magenta dash dot dot line is with $\hat{c}_{s}^{2}=0$. In between
the sound speed is increasing from top to down with
$\hat{c}_{s}^{2}=0.01, 0.05, 0.2$. On the right two panels, the
top Black solid line is for the sound speed of
$\hat{c}_{s}^{2}=1.0$ and the bottom Magenta dash dot dot line is
with $\hat{c}_{s}^{2}=0$. In between the sound speed is decreasing
from top to down with $\hat{c}_{s}^{2}=0.2, 0.05,
0.01$.\label{fig1}}
\end{center}
\end{figure}

In Fig.\ref{fig1} we illustrate how the CMB temperature
anisotropies of four different DE models charactrized by different
equations of state $w$ change on large scales, for different
constant sound speed $\hat{c}_{s}^{2}$. Recently we know that our
Universe is dominated by the dark energy component. This means the
sound speed of DE can only affect the CMB power spectrum at the
very large observable scale via the late Integrated-Sachs-Wolfe
(ISW) effect which describes the perturbations induced by the
passage of CMB photons through the time evolving gravitational
potential wells during DE domination
\cite{Weller:2003hw,Bean:2003fb,hannestad,Koivisto:2005mm}. The
explicit effect on CMB power spectra caused by the sound speed of
DE appears at the large angular scale $l<20$. However, as the
behavior of the equation of state is close to the Cosmological
Constant boundary $w=-1$, these differences get smaller and even
disappear. The different effects of the two DE models in the
bottom panels are much clearer than the ones in the top panels
which are closer to the boundary.

\section{III. Method and datasets}

During our calculation we choose the commonly used parametrization
of the DE equation of state as \cite{Linderpara}:
\begin{equation}
\label{EOS} w_\DE(a) = w_{0} + w_{1}(1-a)~,
\end{equation}
where $a=1/(1+z)$ is the scale factor and $w_{1}=-dw/da$
characterizes the ``running" of the equation of state. We modify
the current publicly available codes like CMBFAST \cite{cmbfast}
and CAMB \cite{camb} to
allow the DE equation of state to cross the boundary $w=-1$. 
For the parametrization of the equation of state which gets across
-1, we introduce a small positive constant $\epsilon$ to divide
the full range of the allowed value of  $w$ into three parts: 1) $
w > -1 + \epsilon$; 2) $-1 + \epsilon \geq w  \geq-1 - \epsilon$;
and 3) $w < -1 -\epsilon $. We neglect the entropy perturbation
contributions, for the regions 1) and 3) the equation of state
does not get across $-1$ and perturbations are well defined by
solving Eqs.(\ref{dotdelta},\ref{dottheta}). For the case 2), the
density perturbation $\delta$ and velocity perturbation $\theta$,
and the derivatives of $\delta$ and $\theta$ are finite and
continuous for the realistic Quintom Dark Energy models. However,
for the perturbations of the parameterized Quintom there is
clearly a divergence. In our study for such a regime, we match the
perturbation in region 2) to the regions 1) and 3) at the boundary
and set:
\begin{equation}\label{dotx}
\delta'=0 ~~,~~\theta'=0~~.
\end{equation}
In our numerical calculations we have limited the range to be $|
\epsilon |<10^{-5}$ and we find our method is a very good
approximation to the multi-field Quintom DE model. For more
details of this method we refer the readers to our previous
companion papers \cite{Zhao:2005vj,Xia:2005ge}.

In this study we have implemented the publicly available Markov
Chain Monte Carlo package CosmoMC \cite{CosmoMC}, which has been
modified to allow for the inclusion of DE perturbation with the
equation of state getting across $-1$. We assume purely adiabatic
initial conditions and a flat Universe. Our most general parameter
space is:
\begin{equation}
\label{parameter} {\bf P} \equiv (\omega_{b}, \omega_{c},
\Theta_{s}, \tau, \hat{c}_{s}^2, w_{0}, w_{1}, n_{s},
\log[10^{10}A_{s}])
\end{equation}
where $\omega_{b}\equiv\Omega_{b}h^{2}$ and
$\omega_{c}\equiv\Omega_{c}h^{2}$ are the physical baryon and Cold
Dark Matter densities relative to the critical density,
$\Theta_{s}$ is the ratio (multiplied by 100) of the sound horizon
to the angular diameter distance at decoupling, $\tau$ is the
optical depth to re-ionization, $\hat{c}_{s}^2$ is the effective
sound speed of Dark Energy, $A_{s}$ and $n_{s}$ characterize the
primordial scalar power spectrum. For the pivot of the primordial
spectrum we set $k_{s0}=0.05$Mpc$^{-1}$. Furthermore, we make use
of the Hubble Space Telescope (HST) measurement of the Hubble
parameter $H_{0}\equiv 100$h~km~s$^{-1}$~Mpc$^{-1}$ \cite{Hubble}
by multiplying the likelihood by a Gaussian likelihood function
centered around $h=0.72$ and with a standard deviation
$\sigma=0.08$. We also impose a weak Gaussian prior on the baryon
density $\Omega_{b}h^{2}=0.022\pm0.002$ (1 $\sigma$) from Big Bang
Nucleosynthesis \cite{BBN}. Simultaneously we will also use a
cosmic age tophat prior as 10 Gyr $< t_0 <$ 20 Gyr.

In our calculations we have taken the total likelihood to be the
products of the separate likelihoods of CMB, LSS and SNIa. In the
computation of CMB we have included the three-year WMAP (WMAP3)
data with the routine for computing the likelihood supplied by the
WMAP team \cite{wmap3:2006}. For LSS information, we have used the
3D power spectrum of galaxies from the SDSS \cite{Tegmark:2003uf}
and 2dFGRS \cite{Cole:2005sx}. To be conservative but more robust,
in the fittings to the 3D power spectrum of galaxies from the
SDSS, we have used the first 14 bins only, $0.015<k_{\rm
eff}<0.1$, which are supposed to be well within the linear regime
\cite{Tegmark:2003ud}. In the calculation of the likelihood from
SNIa we have marginalized over the nuisance parameter
\cite{DiPietro:2002cz}. The supernova data we used are the
``gold'' set of 157 SNIa published by Riess $et$ $al$ in
Ref.\cite{Riess04}.

For each regular calculation, we run 8 independent chains
comprising of $150,000-300,000$ chain elements and spend thousands
of CPU hours to calculate on a supercomputer. The average
acceptance rate is about $40\%$. We test the convergence of the
chains by Gelman and Rubin criteria \cite{R-1} and find that $R-1$
is of order $0.01$ which is more conservative than the recommended
value $R-1<0.1$.

\begin{figure}[htbp]
\begin{center}
\includegraphics[scale=0.6]{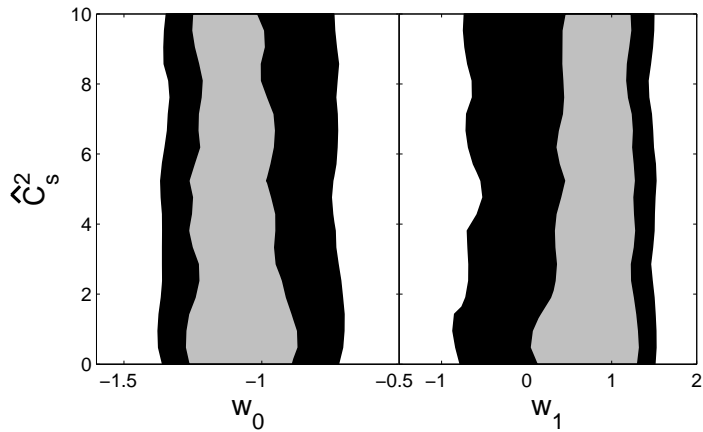}
\caption{Constraints in the ($w_0$, $\hat{c}_{s}^2$) and ($w_1$,
$\hat{c}_{s}^2$) planes at $68\%$ (dark) and $95\%$ (light) C.L.
from a combined analysis of CMB, LSS and SNIa observational data
together.\label{fig2}}
\end{center}
\end{figure}

\section{IV. Global Fitting Results}

In this section we mainly present our global fitting results of
the dark energy parameters $w_0$, $w_1$ and the effective sound
speed of dark energy $\hat{c}_{s}^2$. Firstly, in Fig.\ref{fig2}
we show the constraints on the effective sound speed
$\hat{c}_{s}^2$ from a combined analysis of the CMB, LSS and SNIa
observational data. The likelihood contours of ($w_0$,
$\hat{c}_{s}^2$) and ($w_1$, $\hat{c}_{s}^2$) at $68\%$ and $95\%$
C.L. are nearly vertical lines. The dark energy parameters, as
well as other cosmological parameters which we do not list here,
are almost independent of the effective sound speed of DE. There
is nearly no constraint on the effective sound speed from the
present astronomical data, namely, the current observations are
still not sensitive to $\hat{c}_{s}^2$. Thus the futuristic
precision measurements of CMB on a very large angular scale (low
multipoles) are necessary.

\begin{figure}[htbp]
\begin{center}
\includegraphics[scale=0.4]{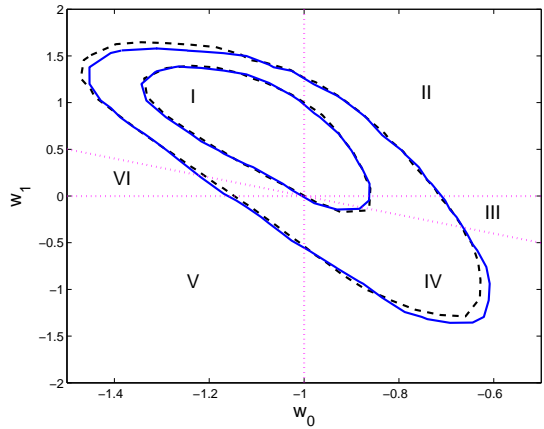}
\caption{Two-dimensional constraints on the DE parameters $w_0$
and $w_1$ from a combined analysis of CMB, LSS and SNIa data at
$68\%$ and $95\%$ C.L. with $\hat{c}_{s}^2=1$ (Black Dashed lines)
and $\hat{c}_{s}^2$ varying (Blue Solid lines). The Magenta Dotted
lines stand for $w_0=-1$ and $w_0+w_1=-1$. The Roman letters
denote different dark energy models. See text for details.
\label{fig3}}
\end{center}
\end{figure}

On the other hand, we consider the effect of the effective sound
speed on the constraints of DE parameters. In Fig.\ref{fig3} we
compare the constraints on the dark energy parameters $w_0$, $w_1$
with $\hat{c}_{s}^2=1$ (Black Dashed lines) and with arbitrary
$\hat{c}_{s}^2$ (Blue Solid lines). As expected, the constraints
on $w_0$, $w_1$ almost unchange. For the model with
$\hat{c}_{s}^2=1$, we find that
$w_0=-1.05_{-0.15-0.30}^{+0.16+0.35}$ and
$w_1=0.53_{-0.55-1.54}^{+0.53+0.75}$, meanwhile,
$w_0=-1.06_{-0.16-0.28}^{+0.17+0.37}$ and
$w_1=0.54_{-0.55-1.52}^{+0.52+0.77}$ for the $\hat{c}_{s}^2$
varying models. We cut the parameter space of $w_{0}-w_{1}$ plane
into six parts by the line of $w_{0}=-1, w_{0}+w_{1}=-1$ and
$w_1=0$. Part III is for quintessence-like models, namely, the
equation of state remains greater than $-1$ regardless of cosmic
time, say, $w>-1$ for past, present and future. Correspondingly,
part VI is for Phantom-like models. Part I,II,V and IV are all for
Quintom-like models. For the models lie within part I and IV,
their equations of state have crossed over -1 till now while the
EoS of the DE models in part II and V will cross -1 in future.

\section{V. Summary}

In this paper we have studied the features of the sound speed of
DE in detail and have used the present observational data to
constrain the effective sound speed of dark energy $\hat{c}_{s}^2$
and the equations of state $w_0$, $w_1$.

If the sound speed is smaller than zero the system is unstable,
due to the divergent classical perturbations. If we regard the DE
models as single barotropic fluid or non-barotropic fluid, when
the equation of state $w$ is close to the boundary $w=-1$ and
$w'|_{w=-1}\neq0$, the classical perturbation of the DE system
will be divergent. In order to settle this problem it is necessary
to add more degrees of freedom. As an example, we analyze the
stability of Quintom DE models with two components. We assume that
the two components have the same sound speed to avoid the
intuitive divergence.

Using the Markov Chain Monte Carlo method, we preform a global
analysis of the $\hat{c}_{s}^2$ and $w_0$, $w_1$. We find that the
current astronomical data have nearly nothing to do with
constraining the  effective sound speed of Dark Energy system. The
constraint on the sound speed of DE is very weak. The futuristic
precision measurements of CMB on a very large angular scale (low
multipoles) are necessary.

\section{Acknowledgements}
We acknowledge the use of the Legacy Archive for Microwave
Background Data Analysis (LAMBDA). The support for LAMBDA is
provided by the NASA Office of Space Science. We have performed
our numerical analysis on the Shanghai Supercomputer Center (SSC).
We would like to thank Mingzhe Li, Hong Li and Pei-Hong Gu for
useful discussions. This work is supported in part by National
Natural Science Foundation of China under Grant Nos. 90303004,
10533010 and 10675136 and by the Chinese Academy of Science under
Grant No. KJCX3-SYW-N2.

\section{Appendix: The Proof of the No-Go Theorem}

In this appendix we present the detailed proof of the ``No-Go"
Theorem which forbids the equation of state parameter of a single
perfect fluid or a single scalar field to cross the $-1$ boundary.

{\bf Theorem}: For theory of dark energy (DE) in the
Friedmann-Robertson-Walker (FRW) universe described by a single
perfect fluid or a single scalar field $\phi$ with a lagrangian of
${\cal L}={\cal L}(\phi, \partial_{\mu}\phi\partial^{\mu}\phi)$,
which minimally couples to Einstein Gravity, its equation of state
$w$ cannot cross over the cosmological constant boundary.

{\bf Proof}: Let us consider the case of fluid firstly. Generally,
a perfect fluid, without viscosity and cannot conduct heat, can be
described by parameters such as pressure $p$, density $\rho$ and
entropy $S$, satisfying the equation of state $p=p(\rho, S)$.
According to the properties of fluid, single perfect fluid can be
classified into two kinds of form, dubbed as barotropic and
non-barotropic.

If the fluid is barotropic, the iso-pressure surface is identical
with the iso-density surface, thus the pressure only depends on
its density in form of $p=p(\rho)$. From Eq.(\ref{adisoseq}), we
can see that the sound speed of a single perfect fluid is
apparently divergent when $w$ crosses $-1$, which leads to
instability in Dark Energy perturbation.

If the fluid is non-barotropic, the pressure generally depends
both on its density and entropy, $p=p(\rho, S)$. The simple form
of the sound speed defined in Eq.(\ref{adisoseq}) is not
well-defined. From Eq.(\ref{cs2}) while taking gravitational gauge
invariance into consideration, we can obtain a more general
relationship between the pressure and the energy density as
follows,\be\label{gauge sound speed} \delta\hat p=\hat
c_s^2\delta\hat\rho~,\ee where the hat denotes the gauge
invariance. Eq.(\ref{Deltaeq}) can be written in a more explicit
form as:
\begin{equation}\label{gaugerho}
\delta\hat\rho=\delta\rho+3{\cal H}(\bar\rho+\bar
p)\frac{\theta}{k^2}~,\end{equation} and correspondingly,
gauge-invariant perturbation of pressure turns out to be
\begin{equation}\label{gaugep}
\delta\hat p=\delta p+3{\cal H}c_a^2(\bar\rho+\bar
p)\frac{\theta}{k^2}~,
\end{equation}
where $\bar\rho$ and $\bar p$ are background energy density and
pressure respectively, while $\theta$ is the perturbation of
velocity as defined in the main text. The gauge-invariant
intrinsic entropy perturbation $\Gamma$ can be described as,
\begin{eqnarray}
\Gamma=\frac{1}{w\bar\rho}(\delta
p-c_a^2\delta\rho)=\frac{1}{w\bar\rho}(\delta\hat
p-c_a^2\delta\hat\rho)~.
\end{eqnarray}
Combining Eqs.(\ref{gauge sound speed}-\ref{gaugep}), we obtain
the following expression,\label{deltap}\bea \delta
p&=&\hat{c_s^2}\delta\rho+\frac{3{\cal
H}\bar\rho\theta(1+w)(\hat{c_s^2}-c_a^2)}{k^2}\nonumber\\
&=&\hat{c_s^2}\delta\rho+\hat{c_s^2}\frac{3{\cal
H}\bar\rho\theta(1+w)}{k^2}-\frac{3{\cal
H}\bar\rho\theta(1+w)}{k^2}w+\frac{\bar\rho\theta w'}{k^2}~.\eea
From the Dark Energy perturbation equation Eq.(\ref{dottheta}),
one can see that $\theta$ will be divergent when $w$ crosses $-1$,
unless $\theta$ satisfies the condition\be\label{condition} \theta
w'=k^2\frac{\delta p}{\bar\rho}~.\ee So we check what will occur
when the condition is satisfied.

By substituting the definition of adiabatic sound speed $c^2_a$
and the condition Eq.(\ref{condition}) into Eq.(\ref{gaugep}), we
obtain $\delta\hat p=0$. Note that
$\Gamma=\frac{1}{w\bar\rho}(\delta\hat p-c_a^2\delta\hat\rho)$, it
is obvious that due to the divergence of $c_a^2$ at the crossing
point, we have to require $\delta\hat\rho=0$ to maintain a finite
$\Gamma$. So we come to the last possibility, that is, $\delta\hat
p=0$ and $\delta\hat\rho=0$. From
Eqs.(\ref{gaugerho},\ref{gaugep}), this case requires that $\delta
p=-c_a^2\frac{3{\cal H}\bar\rho\theta(1+w)}{k^2}$ and
$\delta\rho=-\frac{3{\cal H}\bar\rho\theta(1+w)}{k^2}$, and thus
$\delta p=c_a^2\delta\rho$. It returns to the case of adiabatic
perturbation, which is divergent as mentioned above.

Finally, from the analysis of classical stability, we demonstrate
that there is no possibility for a single perfect fluid to realize
$w$ crossing $-1$. For other proofs, see
\cite{Hu:2004kh,Kunz:2006wc}.

In the following part we will discuss the case of a single scalar
field. The analysis is an extension of the discussion in
\cite{Zhao:2005vj}. The action of the field is given by \be S=\int
d^4 x \sqrt{-g}{\cal L}(\phi,
\partial_{\mu}\phi\partial^{\mu}\phi)~,\ee
where $g$ is the determinant of the metric $g_{\mu\nu}$. To study
the equation of state (EOS) of DE, we firstly write down its
energy-momentum tensor. By definition that
$\delta_{g_{\mu\nu}}S=-\int d^4 x
\frac{\sqrt{-g}}{2}T^{\mu\nu}\delta g_{\mu\nu}$, one can get \be
T^{\mu\nu}=-g^{\mu\nu}{\cal L}-2 \frac{\delta{\cal L}}{\delta
g_{\mu\nu}}=-g^{\mu\nu}{\cal L}+\frac{\partial {\cal L}}{\partial
X}\partial^{\mu}\phi\partial^{\nu}\phi~,\ee where $X \equiv
\frac{1}{2}\partial_{\mu}\phi\partial^{\mu}\phi$. In FRW universe,
we have the metric in form of $g_{\mu\nu}=diag(1, -a^2, -a^2,
-a^2)$. In the framework of Einstein Gravity, we can neglect the
spatial derivatives of $\phi$ and rewrite $X$ as
$X=\frac{1}{2}\dot\phi^2$, where the overdot denotes the
derivative with respect to cosmic time $t$. So $T^{\mu\nu}$ has
the matrix form of $diag(\rho, \frac{p}{a^2}, \frac{p}{a^2},
\frac{p}{a^2})$, comparing with the fluid definition:
$T^{\mu\nu}=(\rho+p)U^\mu U^\nu-p g^{\mu\nu}$ (or
$T^{\mu}_{\nu}=(\rho+p)U^{\mu}U_{\nu}-p\delta^{\mu}_{\nu}$) where
$U^\mu$ is the four-velocity of the fluid. Thus one can get:
\bea p&=&-T^{i}_{i}={\cal L}~,\\
\rho&=& T^{0}_{0}=-\delta^{0}_{0}{\cal L}+\frac{\partial{\cal
L}}{\partial
X}\partial^{0}\phi\partial_{0}\phi~,\nonumber\\
&=&2 X p_{,X}-p~,\eea where $``_{,X}"$ stands for
$``\frac{\partial}{\partial X}"$.

Using the formulae above, the equation of state $w$ is given by:
\be w=\frac{p}{\rho}=\frac{p}{2 X
p_{,X}-p}=-1+\frac{2Xp_{,X}}{2Xp_{,X}-p}~.\ee This means that, at
the crossing point $t^*$, $Xp_{,X}|_{t^*}=0$ must vanish. Since
$w$ needs to cross $-1$, it is required that $Xp_{,X}$ changes
sign before and after the crossing point. That is, in the
neighborhood of $t^*$, $(t^*-\epsilon,t^*+\epsilon)$, we have
\begin{eqnarray}\label{crosscondition}
Xp_{,X}|_{t^*-\epsilon}\cdot Xp_{,X}|_{t^*+\epsilon}<0~.
\end{eqnarray}
Since $X=\frac{1}{2}\dot\phi^2$ is non-negative, the Eq.
(\ref{crosscondition}) can be simplified as
$p_{,X}|_{t^*-\epsilon}\cdot p_{,X}|_{t^*+\epsilon}<0$. Due to the
continuity of perturbation during the crossing epoch, we obtain
$p_{,X}|_{t^*}=0$.

Here we consider the perturbation of the field. We calculate the
perturbation equation with respect to conformal time $\eta$ in
form of: \be u''-c_s^2\nabla^2 u-[\frac{z''}{z}+3c_s^2({\cal
H}'-{\cal H}^2)]u=0~,\ee where we define \be u\equiv
az\frac{\delta\phi}{\phi'}~,~~~~~~~z\equiv
\sqrt{\phi'^2|\rho_{,X}|}~,\ee and prime denotes
$``\frac{d}{d\eta}"$ and $``{\cal H}=\frac{da}{ad\eta}"$.

When expanding the perturbation function $u$ by Fourier
transformation, one can easily obtain the dispersion relation: \be
\omega^2=c_s^2k^2-\frac{z''}{z}-3c_s^2({\cal H}'-{\cal H}^2)~,\ee
with $c_s^2$ defined as $\frac{p_{,X}}{\rho_{,X}}$. To make the
system stable, we need $c_s^2>0$. Note that at the crossing point
we require that $p_{,X}=0$ and $p_{,X}$ will change its sign
during crossing, one can always find a small region where
$c_s^2<0$ unless $\rho_{,X}$ also becomes zero at the crossing
point with the similar behavior of $p_{,X}$. Therefore, the
parameter $z$ will vanish when crossing.

At the crossing point, ${\cal H}'-{\cal H}^2$ in the last term in
the expression of $\omega^2$ is finite, and when assuming that the
universe is fulfilled by that scalar field, it turns out to be
zero.

Since at the crossing point, we have $\rho_{,X}=0$, $z=0$.
Consequently, if $z''\neq0$ at the crossing point the term
$\frac{z''}{z}$ will be divergent. Note that, even if $z''=0$ this
conclusion is still valid. In that case,
$\frac{z''}{z}=\frac{z^{'''}}{z^{'}}$ due to the L'Hospital
theorem. Since $z$ is a non-negative parameter, $z=0$ is its
minimum and at that point $z'$ must vanish,
$\frac{z^{'''}}{z^{'}}$ is either divergent or equal to
$\frac{z^{(4)}}{z''}$ where $z''$ is also equal to zero as we
discussed before. Along this way, if we assume the first
$(n-1)$-th derivative of $z$ with respect to $\eta$ vanishes at
the crossing point and $z^{(n)}\neq 0$, which can be applied to an
arbitrary positive integer $n$, we can always use the L'Hospital
theorem until we find that
$\frac{z''}{z}=\frac{z^{(n)}}{z^{(n-2)}}$, which will still be
divergent. Therefore, the dispersion relation will be divergent at
the crossing point as well, and hence the perturbation will also
not be stable.

In summary, we have analyzed the most general case of a single
scalar field described by a lagrangian in form of ${\cal L}={\cal
L}(\phi, \partial_{\mu}\phi\partial^{\mu}\phi)$, and have studied
different possibilities of $w$ crossing the cosmological constant
boundary. We showed that those cases can either cause the
effective sound speed $c_s^2$ to be negative, or lead to a
divergent dispersion relation, which makes the system unstable.

To conclude, we have proved that in FRW universe, it is impossible
for a single perfect fluid or a single scalar field minimally
coupled to Einstein Gravity to have its equation of state $w$
crossing the cosmological constant boundary.

To realize $w$ across $-1$, one must introduce extra degrees of
freedom or introduce the nonminimal couplings or modify the
Einstein gravity. In the recent years there have been a lot of
activities in building models with $w$ crossing $-1$
\cite{Cai:2005ie,Apostolopoulos:2006si,Zhang:2006at}.

The simplest Quintom model is to introduce two scalar fields with
one being quintessence-like and the other phantom-like
\cite{Feng:2004ad,Zhang:2005eg,Guo:2004fq}. However, this model
suffers from the problem of quantum instability which is inherited
from phantom \cite{quantum1,Kahya:2006hc,quantum2,Hawking:2001yt}.
This issue could be solved in the effective description of the
quintom model with the operator $\phi\Box\phi$ involved. Consider
a canonical scalar field with the lagrangian ${\cal L} =
\frac{1}{2}
\partial_\mu \phi\partial^\mu \phi - V( \phi )$. This model is well defined
at quantum level, however, it does not give $w$ crossing $-1$. As
an effective theory as we know the lagrangian should include more
operators. If these operators are functions of only the scalar
field $\phi$ and its first derivative $\partial_\mu \phi
\partial^\mu \phi$, as we proved above the $w$ still can not cross
over -1. However as pointed out in
Refs.\cite{Li:2005fm,Zhang:2006ck}, the existence of the operator
$\phi\Box\phi$ makes it possible for $w$ to cross over the
cosmological constant boundary. Furthermore, at quantum level as
an perturbation theory this effective model is well defined. The
connection of this type of Quintom-like theory to the string
theory has been considered in \cite{tachyon} and
\cite{Aref'eva:2006et}.

\end{document}